\documentclass[
 aps,
 ,onecolumn
]{revtex4}
\usepackage{graphicx}
\usepackage{amsmath}
\usepackage{mathrsfs}

\usepackage{tikz}
\usepackage{pgflibraryarrows}
\usepackage{pgflibrarysnakes}
\usetikzlibrary{decorations.pathmorphing}
\usepgflibrary{arrows.meta}

\begin{document}

\title{Decoherence of the radiation from an accelerated quantum source}

\author{Daiqin Su}
\email{sudaiqin@gmail.com}
\author{T.~C.~Ralph}\email{ralph@physics.uq.edu.au}
\affiliation{Centre for Quantum Computation and Communication Technology, School of Mathematics and Physics, 
The University of Queensland, St. Lucia, Queensland, 4072, Australia}



\date{\today}


\begin{abstract}
{
Decoherence is the process via which quantum superpositions states are reduced to classical mixtures. Decoherence has been predicted for relativistically accelerated quantum systems, however examples to date have involved restricting the detected field modes to particular regions of space-time. If the global state over all space-time is measured then unitarity returns and the decoherence is removed. Here we study a decoherence effect associated with accelerated systems that cannot be explained in this way. In particular we study a uniformly accelerated source of a quantum field state - a single-mode squeezer. Even though the initial state of the field is vacuum (a pure state) and the interaction with the quantum source in the accelerated frame is unitary, we find that the final state detected by inertial observers is decohered, i.e. in a mixed state. This unexpected result may indicate new directions in resolving inconsistencies between relativity and quantum theory. We extend this result to a two-mode state and find entanglement is also decohered. 

}
\end{abstract}

\maketitle



Unitary evolution is one of the fundamental assumptions of quantum mechanics. An initial pure state of an isolated quantum system always evolves into another pure state. The situation is not as simple when we consider non-inertial, relativistic frames of reference. For example, the transformation between the description of the quantum vacuum state as seen by inertial observers and the description of the same state by uniformly accelerated observers is not strictly unitary. Never-the-less it is still assumed that in transforming between reference frames pure states will always evolve to pure states provided that the entire space-time is included.

Consider an inertial observer who constantly observes a massless field prepared in the Minkowski vacuum state. By definition they will observe no particles. However, according to the Unruh/Davies effect \cite{Unruh76, Davies75}, a uniformly accelerating observer who constantly observes the same field will see thermal radiation (Unruh radiation), and hence will count particles. The vacuum state is pure whilst a thermal state is mixed, seemingly implying a non-unitary evolution. The resolution is that a single accelerating observer is restricted to a section of space-time called a Rindler wedge. By introducing a second, mirror image accelerated observer we find that the thermal state can be purified into a two-mode squeezed state  \cite{Unruh84, Lee86, Takagi86} and unitarity is restored. 

Because of the equivalence principle there is a strong relationship between gravity and acceleration \cite{MTW}. The analogous situation to Unruh radiation in curved space-time is that of thermal radiation from black holes (Hawking radiation) \cite{Hawking75}. In this case regaining unitarity is not straightforward because the analogue of the mirror image Rindler wedge lies behind the black hole event horizon and so is inaccessible.  Given that in the far future the black hole is expected to completely evaporate, this leads to the black hole information paradox \cite{Hawking77}. In spite of many attempts \cite{Susskind93, Stephens94, Mathur05, Hawking16, Baccetti16}, a completely satisfactory resolution of this problem has not been found \cite{AMPS13, Braunstein13}.

%

In this paper we consider accelerated quantum systems in flat space, however we set up the problem differently such that we explicitly start and end with global, inertial observers. In the intermediate region we allow interactions with an accelerated system. Unexpectedly we find a decoherence effect that only affects non-classical quantum states and appears even though the observers have access to the entire space-time.

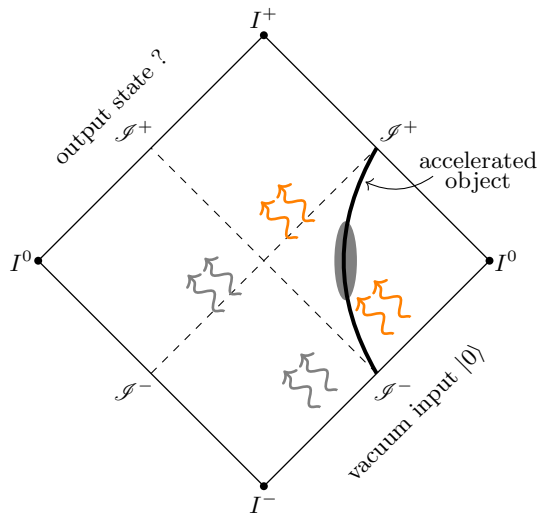
\begin{figure}[ht!]

\begin{tikzpicture}[scale=0.75]

\draw [line width=0.5pt] (4, 0) -- (0, 4) -- (-4, 0) -- (0, -4) -- cycle;

\begin{scope}
\fill (4, 0) circle (2pt);
\fill (0, 4) circle (2pt);
\fill (-4, 0) circle (2pt);
\fill (0, -4) circle (2pt);
\end{scope}

\begin{scope}
\draw (4.3, 0) node {$I^0$};
\draw (-4.3, 0) node {$I^0$};
\draw (0, 4.3) node {$I^+$};
\draw (0, -4.3) node {$I^-$};
\draw (2.3, -2.3) node {$\mathscr{I}^{-}$};
\draw (-2.3, -2.3) node {$\mathscr{I}^{-}$};
\draw (2.4, 2.3) node {$\mathscr{I}^{+}$};
\draw (-2.3, 2.3) node {$\mathscr{I}^{+}$};
\end{scope}

\begin{scope} [dashed]
  \draw (-2, -2) -- (2, 2);
  \draw (2, -2) -- (-2, 2);
\end{scope}

\fill [color=gray] (1.45, 0) circle (0.2 and 0.7);
\draw [line width=1.5pt] (2, -2) edge [bend left] (2, 2); 

\begin{scope} [color=orange, line width=1pt]
\draw [->, decorate, decoration=snake] (2.3, -1.3) -- (1.65, -0.65);
\draw [->, decorate, decoration=snake] (2.6, -1.) -- (1.95, -0.35);
\draw [->, decorate, decoration=snake] (0.6, 0.4) -- (-0.05, 1.05);
\draw [->, decorate, decoration=snake] (0.9, 0.7) -- (0.25, 1.35);
\end{scope}

\begin{scope} [color=gray, line width=1pt]
\draw [->, decorate, decoration=snake] (1.3, -2.3) -- (0.65, -1.65);
\draw [->, decorate, decoration=snake] (1., -2.6) -- (0.35, -1.95);
\draw [->, decorate, decoration=snake] (-0.4, -0.6) -- (-1.05, 0.05);
\draw [->, decorate, decoration=snake] (-0.7, -0.9) -- (-1.35, -0.25);
\end{scope}

\begin{scope}
\draw (2.7, -2.7) node [rotate=45] {vacuum input $| 0 \rangle$};
\draw (-2.7, 2.7) node [rotate=45] {output state ? };
\draw (3.8, 1.8) node [rotate=0] {accelerated};
\draw (3.8, 1.4) node [rotate=0] {object};
\end{scope}

\path [->] (3, 1.5) edge [bend left] (1.8, 1.4);

\end{tikzpicture}

\caption{\footnotesize Penrose diagram of Minkowski spacetime. $I^0$ is the spatial infinity, $I^-$ and $I^+$ are the 
past and future infinities, $\mathscr{I}^-$ and $\mathscr{I}^+$ are the past and future null infinities. A uniformly accelerated object follows the
black worldline. Interactions between the accelerated object and the field are localized in Rindler time, represented by the shaded region.  } 
\label{PenroseDiagram}
\end{figure}

The specific problem we will analyse is summarized by the Penrose diagram \cite{MTW} in
Fig. \ref{PenroseDiagram}. An object uniformly accelerates in the right Rindler wedge (black curve). Interactions with a massless scalar field are unitarily turned on 
and off during its lifetime (shaded region) such that it interacts with a single spatiotemporal mode in the accelerated (Rindler) coordinates. In the past null infinity $\mathscr{I}^-$, the initial state of the field is set to be the Minkowski vacuum. 
For simplicity we consider a 1+1 theory in which the right and left moving fields are decoupled. We assume the right moving field modes are unaffected by the accelerating object. The output state of the left moving field modes in the future null infinity $\mathscr{I}^+$ is detected by inertial, Minkowski detectors. We ask whether the detected field is in a pure state.
We extend this picture to two-mode sources that entangle right and left movers later in the paper. 

%

\vspace{0.5cm}
\noindent \textbf{Detection of the state} 

\noindent The Minkowski detectors are modelled by the Hermitian number operators, $\hat N_k = \hat a_k^{\dagger} \hat a_k$, where $\hat a_k$ ($\hat a_k^{\dag}$) are the Minkowski field annihilation (creation) operators for wave-number $k$. The frequencies $\Omega = |k|$ are with respect to the proper time of the inertial reference frame under consideration (note we are using units for which $c = 1$). The excitation probability of an ideal, inertial, 2-level system of resonant frequency $\Omega$, coupled weakly to the field, is proportional to $\langle \hat N_k \rangle$ \cite{Scully97}. We can model a finite bandwidth detector via the operator $\hat{N}_{\Delta k} = \int_{k_o - \Delta k}^{k_o + \Delta k} \mathrm{d} k \; \hat{a}^{\dag}_{k}\hat{a}_{k}$. If the bandwidth of the detector is much larger than that of the mode under consideration then we can extend the limits of integration to $\pm \infty$ and so define $\hat{N} = \int \mathrm{d}k \; \hat{a}^{\dag}_{k}\hat{a}_{k}$. Note that by definition $\langle 0| \hat N |0 \rangle = 0$ for the Minkowski vacuum state, $| 0 \rangle$. 

In order to characterize the state of a particular field mode we use homodyne tomography \cite{Lvovsky09}. 
In homodyne tomography, the Wigner function \cite{Scully97} of the state is reconstructed from measurements of the moments of quadrature amplitudes via homodyne detection. For Gaussian states it is sufficient to measure and analyse only the first and second order moments \cite{Weedbrook12}.
In homodyne detection \cite{BachorRalph}, a weak signal field and a strong local oscillator are 
coherently combined and measured with broad-band detection as discussed above. For simplicity and to stay within the 1+1, scalar field paradigm, we specifically use self-homodyne detection here. In self-homodyne detection, the signal field is displaced by a strong local oscillator directly, and the output field is detected. 
Assume that the signal field mode operator is $\hat{a} = \int \mathrm{d} k f(k) \hat{a}_{k}$ and the local oscillator is a strong coherent state $| \alpha \rangle$, prepared in the same field mode (characterized by $f(k)$) with $\alpha$ a complex number,
$\alpha = |\alpha| e^{i\phi}$, and $|\alpha| \gg 1$. The photon number operator can be shown to be
\begin{equation}
\hat{N}(\phi) \approx |\alpha|^2 + |\alpha| \hat{X}(\phi) 
\end{equation}
where $\hat{X}(\phi) = \hat{a} e^{-i \phi} + \hat{a}^{\dag} e^{i\phi}$ is the quadrature amplitude of the signal field and a term not multiplied by $|\alpha|$ has been neglected as small. As a reference we can also consider the operator
\begin{equation}
\hat{N}_0 \approx |\alpha|^2 + |\alpha| \hat{X}_v
\end{equation}
representing the situation where the signal is not imposed and so $\hat v$ represents the mode when it is prepared in the vacuum state. Hence the average quadrature amplitude of the field is given by
\begin{equation}
\langle \hat{X}(\phi) \rangle = {{\langle \hat{N}(\phi) \rangle - \langle \hat{N}_0 \rangle} \over {\sqrt{\langle \hat{N}_0 \rangle}}}
\label{quad}
\end{equation}
where we have used $\langle \hat{X}_v \rangle =0$. Its variance is given by
\begin{equation}
 \big(\Delta X(\phi) \big)^2 = {{\big(\Delta N(\phi)\big)^2} \over {\langle \hat{N}_0 \rangle}}. 
 \label{var}
\end{equation}
For the Gaussian states considered here this will be sufficient to completely characterize them. 
We wish to apply this technique to the output state from the interactions between a uniformly accelerated object and the scalar field. In order to match the mode
shape of the local oscillator to that of the output signal field, we assume that the local oscillator is also imposed in the accelerated frame in a matching mode to the signal.

\vspace{0.5cm}
\noindent \textbf{Interaction with the accelerated source} 

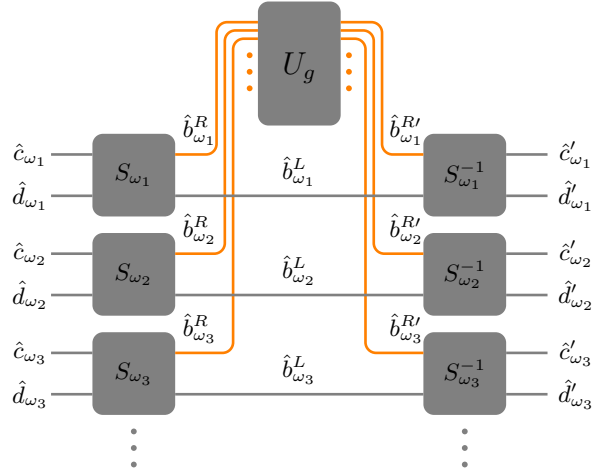
\begin{figure}[ht!]

\begin{tikzpicture}[scale=1.1]

\begin{scope} [rounded corners, color=gray]
\fill (0, 0) rectangle +(1, 1) ;
\fill (0, 1.2) rectangle +(1, 1) ;
\fill (0, 2.4) rectangle +(1, 1) ;

\fill (4, 0) rectangle +(1, 1) ;
\fill (4, 1.2) rectangle +(1, 1) ;
\fill (4, 2.4) rectangle +(1, 1) ;

\fill (2, 3.5) rectangle +(1, 1.5);
\end{scope};
    
\begin{scope} [line width = 1pt, color = orange]
\draw (1, 3.15) [rounded corners] -- (1.5, 3.15) -- (1.5, 4.75) -- (2, 4.75); 
\draw (3, 4.75) [rounded corners] -- (3.5, 4.75) -- (3.5, 3.15) -- (4, 3.15);
\draw (1, 1.95) [rounded corners] -- (1.6, 1.95) -- (1.6, 4.65) -- (2, 4.65); 
\draw (3, 4.65) [rounded corners] -- (3.4, 4.65) -- (3.4, 1.95) -- (4, 1.95);
\draw (1, 0.75) [rounded corners] -- (1.7, 0.75) -- (1.7, 4.55) -- (2, 4.55); 
\draw (3, 4.55) [rounded corners] -- (3.3, 4.55) -- (3.3, 0.75) -- (4, 0.75);
\end{scope};

\begin{scope} [line width = 1pt, color = orange]
\fill (1.9, 4.35) circle (1pt);
\fill (1.9, 4.15) circle (1pt);
\fill (1.9, 3.95) circle (1pt);

\fill (3.1, 4.35) circle (1pt);
\fill (3.1, 4.15) circle (1pt);
\fill (3.1, 3.95) circle (1pt);
\end{scope};

\begin{scope} [line width = 1pt, color = gray]
\draw (-0.5, 0.25) -- (0, 0.25); 
\draw (-0.5, 0.75) -- (0, 0.75);
\draw (-0.5, 1.45) -- (0, 1.45); 
\draw (-0.5, 1.95) -- (0, 1.95);  
\draw (-0.5, 2.65) -- (0, 2.65); 
\draw (-0.5, 3.15) -- (0, 3.15); 
\end{scope};

\begin{scope} [line width = 1pt, color = gray]
\draw (5, 0.25) -- (5.5, 0.25); 
\draw (5, 0.75) -- (5.5, 0.75);
\draw (5, 1.45) -- (5.5, 1.45); 
\draw (5, 1.95) -- (5.5, 1.95);  
\draw (5, 2.65) -- (5.5, 2.65); 
\draw (5, 3.15) -- (5.5, 3.15); 
\end{scope};

\begin{scope} [line width = 1pt, color = gray]
\draw (1, 0.25) -- (4, 0.25); 
\draw (1, 1.45) -- (4, 1.45);   
\draw (1, 2.65) -- (4, 2.65); 
\end{scope};

\draw (2.5, 4.2) node  {\large $U_g$};

\draw (0.5, 0.5) node  {$S_{\omega_3}$};
\draw (0.5, 1.7) node  {$S_{\omega_2}$};
\draw (0.5, 2.9) node  {$S_{\omega_1}$};

\draw (4.5, 0.5) node  {$S^{-1}_{\omega_3}$};
\draw (4.5, 1.7) node  {$S^{-1}_{\omega_2}$};
\draw (4.5, 2.9) node  {$S^{-1}_{\omega_1}$};

\draw (1.3, 1.05) node {$\hat{b}^R_{\omega_3}$};
\draw (1.3, 2.25) node {$\hat{b}^R_{\omega_2}$};
\draw (1.3, 3.45) node {$\hat{b}^R_{\omega_1}$};

\draw (3.8, 1.05) node {$\hat{b}^{R\prime}_{\omega_3}$};
\draw (3.8, 2.25) node {$\hat{b}^{R\prime}_{\omega_2}$};
\draw (3.8, 3.45) node {$\hat{b}^{R\prime}_{\omega_1}$};

\draw (2.5, 0.55) node {$\hat{b}^L_{\omega_3}$};
\draw (2.5, 1.75) node {$\hat{b}^L_{\omega_2}$};
\draw (2.5, 2.95) node {$\hat{b}^L_{\omega_1}$};

\draw (-0.75, 0.25) node {$\hat{d}_{\omega_3}$};
\draw (-0.75, 0.75) node {$\hat{c}_{\omega_3}$};
\draw (-0.75, 1.45) node {$\hat{d}_{\omega_2}$};
\draw (-0.75, 1.95) node {$\hat{c}_{\omega_2}$};
\draw (-0.75, 2.65) node {$\hat{d}_{\omega_1}$};
\draw (-0.75, 3.15) node {$\hat{c}_{\omega_1}$};

\draw (5.85, 0.25) node {$\hat{d}^{\prime}_{\omega_3}$};
\draw (5.85, 0.75) node {$\hat{c}^{\prime}_{\omega_3}$};
\draw (5.85, 1.45) node {$\hat{d}^{\prime}_{\omega_2}$};
\draw (5.85, 1.95) node {$\hat{c}^{\prime}_{\omega_2}$};
\draw (5.85, 2.65) node {$\hat{d}^{\prime}_{\omega_1}$};
\draw (5.85, 3.15) node {$\hat{c}^{\prime}_{\omega_1}$};

\begin{scope} [line width = 1pt, color = gray]
\fill (0.5, -0.2) circle (1pt);
\fill (0.5, -0.4) circle (1pt);
\fill (0.5, -0.6) circle (1pt);

\fill (4.5, -0.2) circle (1pt);
\fill (4.5, -0.4) circle (1pt);
\fill (4.5, -0.6) circle (1pt);
\end{scope};

\end{tikzpicture}

\caption{\footnotesize Circuit for a uniformly accelerated object. Rindler modes in the right Rindler wedge interact
with the object, which is represented by the unitary operator $\hat U_g$, while Rindler modes in the left Rindler wedge remain unaffected. 
The time dependent interactions mix different frequency Rindler modes.} 
\label{GeneralCircuit}
\end{figure}

\noindent Interactions between uniformly accelerated objects (Unruh-DeWitt detectors, mirrors etc.) and quantum fields have been studied for
many years \cite{Birrell84, Unruh92, Parentani95, Obadia01}. Recently, a non-perturbative quantum circuit model was proposed to investigate these interactions and calculate radiation
from a uniformly accelerated object \cite{Su16}.  Here we generalize the circuit model to include time dependent interactions. 
The relevant circuit is shown in Fig. \ref{GeneralCircuit}. The circuit models the interaction as a Heisenberg evolution of Unruh mode operators \cite{Unruh76}, $\hat c_{\omega}$, $\hat d_{\omega}$ to Rindler operators, $\hat b_{\omega}^L$, $\hat b_{\omega}^R$, then back to Unruh operators. The Rindler operators represent the natural modes that uniformly accelerated systems interact with. The frequency, $\omega$, is with respect to the proper time of the accelerated observer. The Unruh operators are a useful mathematical stepping stone between the accelerated and inertial reference frames. The Minkowski modes, $\hat a_k$, that represent our inertial detection scheme are then constructed from the output Unruh modes -- this final step is not represented by a circuit.

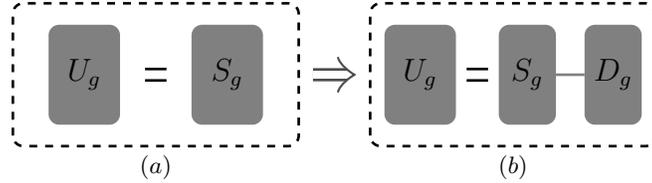
\begin{figure}[ht!]

\begin{tikzpicture}[scale=0.95]

\begin{scope} [dashed, line width = 1pt, rounded corners]
\draw (0, 0) rectangle +(4, 2);
\draw (5, 0) rectangle +(4, 2);
\end{scope};

\begin{scope} [color = gray, rounded corners]
\fill (0.5, 0.3) rectangle +(1, 1.4);
\fill (5.2, 0.3) rectangle +(1, 1.4);

\fill (2.5, 0.3) rectangle +(1, 1.4);
\fill (6.8, 0.3) rectangle +(0.8, 1.4);
\fill (8, 0.3) rectangle +(0.8, 1.4);
\end{scope}

\draw [line width = 1pt] (6.35, 0.92) -- (6.65, 0.92);
\draw [line width = 1pt] (6.35, 1.08) -- (6.65, 1.08);

\draw [line width = 1pt] (1.85, 0.92) -- (2.15, 0.92);
\draw [line width = 1pt] (1.85, 1.08) -- (2.15, 1.08);

\draw [line width = 1pt, color = gray]  (7.6, 1) -- (8, 1);

\draw [color=black!70, line width=1pt, double distance=3pt, -{Classical TikZ Rightarrow[length=2.5mm]}]  (4.2, 1) -- (4.8, 1);

\draw (1, 1) node {\large $U_g$};
\draw (3, 1) node {\large $S_g$};

\draw (5.7, 1) node {\large $U_g$};
\draw (7.2, 1) node {\large $S_g$};
\draw (8.4, 1) node {\large $D_g$};

\draw (2, -0.3) node {$(a)$};
\draw (7, -0.3) node {$(b)$};

\end{tikzpicture}

\caption{\footnotesize Self-homodyne detection. (a) A signal unitary $\hat S_g$ generates quantum signals that we are going to analyse. 
(b) A displacement is added after the signal unitary  $\hat S_g$ to realize homodyne detection. The mode shape of the displacement is perfectly matched to
that of the signal unitary. } 
\label{fig:Homodyne}
\end{figure}

The unitary operator $\hat{U}_g$ acts only on the right Rindler wedge operators, $\hat b_{\omega}^R$, and represents localized interactions between the accelerated object and the scalar 
field. The localization is characterized by the normalized wave packet $g(\omega)$. 
In contrast to the time independent case, 
the time dependent unitary, $\hat{U}_g$, mixes different Rindler frequency modes. The relation between the Rindler modes $\hat{b}^{R\prime}_{\omega}$ and 
$\hat{b}^{R}_{\omega}$ is \cite{Rohde07}
\begin{equation}
\hat{b}^{R\prime}_{\omega} = \hat{b}^{R}_{\omega} + g^*(\omega) \big( \hat{U}^{\dag}_g \hat{b}^R_g \hat{U}_g - \hat{b}^R_g \big),
\end{equation}
where 
$\hat{b}^R_g \equiv \int \mathrm{d} \omega g(\omega) \hat{b}^{R}_{\omega}$ is the localized mode operator satisfying 
commutation relation $[\hat{b}^R_g, \hat{b}^{R\dag}_g]=1$. 
Taking into account the relation between Unruh modes and 
Rinder modes \cite{Su16}, which is basically a two-mode squeezing, we obtain the input-output relations for Unruh modes,  
\begin{eqnarray}\label{GeneralUnruh}
\hat{c}_{\omega}^{\prime}
&=& \hat{c}_{\omega} + g^*(\omega) \cosh r_{\omega}  \big( \hat{U}^{\dag}_g \hat{b}^R_g \hat{U}_g - \hat{b}^R_g \big), \nonumber\\
\hat{d}_{\omega}^{\prime}
&=& \hat{d}_{\omega} -  g(\omega) \sinh r_{\omega} \big( \hat{U}^{\dag}_g \hat{b}^{R\dag}_g \hat{U}_g - \hat{b}^{R\dag}_g \big),
\end{eqnarray}
where the two-mode squeezing factor $r_{\omega}$ is defined as $\tanh r_{\omega} = e^{-\pi \omega/a}$. In equation (\ref{GeneralUnruh}) the 
operator $\hat{b}^R_g$ can be explicitly expressed in terms of the input Unruh modes $\hat{c}_{\omega}$ and $\hat{d}_{\omega}$. 
In the following we will use $\hat{U}_g = \hat{D}_g(\alpha) \hat{S}_g$, where $\hat{S}_g$ creates the quantum signal we wish to analyse, whilst $\hat{D}_g(\alpha) = \exp \big(\alpha \hat{b}^{R\dag}_g - \alpha^* \hat{b}^R_g  \big)$ produces the local oscillator needed for the self-homodyne detection (Fig. \ref{fig:Homodyne}).
It is easy to show that $\hat{D}^{\dag}_g \hat{b}^R_g \hat{D}_g = \hat{b}^R_g + \alpha$ \cite{BachorRalph}.

Finally we require the input-output relations for
Minkowski modes. The transformation from Unruh modes to Minkowski modes is \cite{Su16}
\begin{eqnarray}\label{UnruhMinkowski}
\hat{a}^{\prime}_{k} &=& \int \mathrm{d} \omega ~\big(A_{k\omega} \hat{c}^{\prime}_{\omega} + B_{k\omega} \hat{d}^{\prime}_{\omega} \big),
\end{eqnarray}
where the Bogoliubov transformation coefficients are \cite{Crispino08}
\begin{eqnarray}\label{AB}
A_{k\omega} = B^*_{k\omega} =  \frac{i \sqrt{2 \sinh(\pi \omega/a)}}{2\pi \sqrt{\omega k}} \Gamma(1-i\omega/a) \bigg(\frac{k}{a}\bigg)^{i\omega/a}. 
\end{eqnarray}

The total Minkowski particle number operator is obtained by using equation (\ref{UnruhMinkowski}), 
\begin{eqnarray}\label{ParcleNumber}
\hat{N} 
&=&  \int \mathrm{d}k  \int \mathrm{d} \omega_1 \int \mathrm{d}\omega_2
(A^*_{k \omega_1} \hat{c}^{\prime \dag}_{\omega_1} + B^*_{k \omega_1} \hat{d}^{ \prime \dag}_{\omega_1})
(A_{k \omega_2} \hat{c}^{\prime}_{\omega_2} + B_{k \omega_2} \hat{d}^{\prime}_{\omega_2})\nonumber\\
&=& \int \mathrm{d} \omega ~ (\hat{c}^{\prime \dag}_{\omega}\hat{c}^{ \prime}_{\omega} 
+ \hat{d}^{\prime \dag}_{\omega}\hat{d}^{\prime}_{\omega}),
\end{eqnarray}
where we have used $\int \mathrm{d} k A_{k \omega} A^*_{k \omega'} = \delta(\omega-\omega')$ 
and $\int \mathrm{d} k A_{k \omega} A_{k \omega'} = 0$.
The square of the total particle number operator is 
\begin{eqnarray}\label{SquareNumber}
\hat{N}^2 
&=&  \int \mathrm{d} \omega_1  \int \mathrm{d} \omega_2 ~
\big(\hat{c}^{ \prime \dag}_{\omega_1}\hat{c}^{ \prime}_{\omega_1} \hat{c}^{ \prime \dag}_{\omega_2}\hat{c}^{\prime}_{\omega_2}
+ \hat{d}^{ \prime \dag}_{\omega_1}\hat{d}^{ \prime}_{\omega_1} \hat{d}^{ \prime \dag}_{\omega_2}\hat{d}^{ \prime}_{\omega_2}
+ \hat{c}^{ \prime \dag}_{\omega_1}\hat{c}^{\prime}_{\omega_1}\hat{d}^{ \prime \dag}_{\omega_2}\hat{d}^{ \prime}_{\omega_2}
 +\hat{d}^{ \prime \dag}_{\omega_1}\hat{d}^{\prime}_{\omega_1}\hat{c}^{ \prime \dag}_{\omega_2}\hat{c}^{ \prime}_{\omega_2} \big).
\end{eqnarray}
A full computation of the vacuum expectation value of $\hat{N}^2$ is straightforward but usually tedious. 
However, when the amplitude of displacement is large ($|\alpha| \gg 1$), it is adequate to only keep terms of order $|\alpha|^4$ and $|\alpha|^2$ as per the approximation leading to equations (\ref{quad}) and (\ref{var}).

\vspace{0.5cm}
\noindent \textbf{Classical Signals} 

\noindent We first consider preparing a classical signal on the accelerated mode. In particular, we generate a classical signal by displacing the Rindler mode
$\hat{b}^R_g$ with an amplitude $\beta$. 
This produces a coherent state, the ``most classical" quantum state. The operator that creates this signal is $\hat S_g = \hat D_g(\beta)$, with $|\beta| \ll |\alpha|$. The expectation value and variance of the quadrature amplitudes as observed by the inertial detectors are
\begin{eqnarray}
X_{\beta}(\phi) &=& \sqrt{\mathcal{I}_c + \mathcal{I}_s} \big(\beta e^{-i\phi} + \beta^* e^{i\phi} \big), \nonumber\\
V_{\beta}(\phi) &=& 1,
\label{quadvar}
\end{eqnarray}
where $\mathcal{I}_c = \int \mathrm{d} \omega |g(\omega)|^2 \cosh^2 r_{\omega}$ and $\mathcal{I}_s = \int \mathrm{d} \omega |g(\omega)|^2 \sinh^2 r_{\omega}$. 
Equation (\ref{quadvar}) characterises a pure coherent state.
Therefore, displacing a Rindler mode generates a coherent state with amplitude $(\sqrt{\mathcal{I}_c + \mathcal{I}_s}) \beta$ as viewed by an inertial
observer. As expected the overall evolution is from a pure state to a pure state.

\vspace{0.5cm}
\noindent \textbf{Quantum Signals} 

\noindent A more interesting scenario is that a uniformly accelerated single-mode squeezer squeezes the thermal state in the right Rindler wedge. The single-mode squeezing operator $\hat{S}_{1}(r)$ is defined as \cite{BachorRalph}
\begin{equation}\label{SingleSqueezing}
\hat{S}_{1}(r) =  \exp\bigg\{\frac{r}{2} \big(\hat{b}^{R\dag}_g \big)^2 - \frac{r}{2} \big(\hat{b}^{R}_g \big)^2 \bigg\},
\end{equation}
where $r$ is the squeezing factor and is assumed to be real. The operator that creates quantum signals is $\hat S_g = \hat S_1(r)$ so that the unitary 
$\hat{U}_g = \hat{D}_g(\alpha) \hat{S}_{1}(r)$.
By substituting this unitary into equation (\ref{GeneralUnruh}) one can derive the input-output relations for Unruh modes, 
which are then substituted into equations (\ref{ParcleNumber}) and (\ref{SquareNumber}) to calculate the vacuum expectation value of the Minkowski particle number and the square of the particle number (see Appendix for details). We find that the expectation value of the quadrature amplitude is zero, and the variance is
 \begin{eqnarray}\label{Variance}
 V(\phi) &=& \cosh(2r) + 4  \mathcal{I}_c (\mathcal{I}_c-1) (\cosh 2r -2 \cosh r + 1) 
+ 2 \sinh r \big[(2\mathcal{I}_c -1)^2 \cosh r - 4 \mathcal{I}_c (\mathcal{I}_c-1)\big] \cos(2\phi). 
 \end{eqnarray}
The maximum and minimum variances are obtained when $\phi=0$ and $\phi=\pi/2$, respectively. 
\begin{eqnarray}\label{MaxMinVariance}
V_{\text{max}}
&=& e^{2r} + 4\mathcal{I}_c (\mathcal{I}_c-1) (e^r-1)^2, \nonumber\\
V_{\text{min}}
&=& e^{-2r} + 4\mathcal{I}_c (\mathcal{I}_c-1) (e^{-r}-1)^2. 
 \end{eqnarray}
It is evident from equations (\ref{Variance})  and (\ref{MaxMinVariance}) that noises are added onto the variance of the original single-mode squeezed state. The amount of additional noises depends on the squeezing factor $r$ and $\mathcal{I}_c$.  A question of particular interest is whether the final state is a pure state. For Gaussian states, the criterion for purity is that the product of maximum and minimum variances is unity \cite{BachorRalph}. From equation (\ref{MaxMinVariance}) we find the product of the maximum and minimum variances is
\begin{eqnarray}\label{ProductMaxMin}
V_{\text{max}}V_{\text{min}}
&=& 1 + 16  \mathcal{I}_c (\mathcal{I}_c-1)(\cosh r -1) \cosh r + 64 \mathcal{I}_c^2 (\mathcal{I}_c-1)^2(\cosh r -1)^2. 
\end{eqnarray}
We can see that the product is always greater than one unless $r=0$ or $\mathcal{I}_c = 1$. This is our main result. Unexpectedly, the inertial observer sees a decoherence effect that in general takes the initial pure state to a mixed state.

The case of $r=0$ means the accelerated object does nothing so that the output state is the Minkowski vacuum. 
$\mathcal{I}_c$ 
can be approximated as $\mathcal{I}_c \approx e^{2\pi \omega_0/a}/(e^{2\pi\omega_0/a}-1)$ when $g(\omega)$ is a very narrow bandwidth wave packet with central frequency $\omega_0$.  When $2\pi \omega_0/a \rightarrow \infty$, $\mathcal{I}_c \rightarrow 1$ so that $V_{\text{min}} \rightarrow e^{-2r}$ and $V_{\text{max}} \rightarrow e^{2r}$. This corresponds to a single-mode squeezed vacuum state, which is pure. The above limit could happen in two cases. The first is that the central frequency $\omega_0$ is fixed while $a \rightarrow 0$. This means the single-mode squeezer tends to be static in an inertial frame. It thus produces the standard single-mode squeezed vacuum state. The second case is that $a$ is fixed and finite, while $\omega_0 \rightarrow \infty$. It is well known that a uniformly accelerated observer experiences a thermal radiation with temperature $T_U = \frac{a}{2\pi}$ in the Minkowski vacuum \cite{Unruh76}. The spectral distribution of the thermal radiation follows the Plank's law, which exponentially decays in the high frequency limit. Or equivalently, the high frequency tail of a thermal state looks almost like a vacuum. Therefore the single-mode squeezer that squeezes the high frequency tail of the Unruh radiation produces a squeezed vacuum state. Overall, when the Unruh effect is not significant, a uniformly accelerated single-mode squeezer produces the standard single-mode squeezed vacuum state. Otherwise, the product of the maximum and minimum variances is greater than one, indicating that the output state is mixed. 

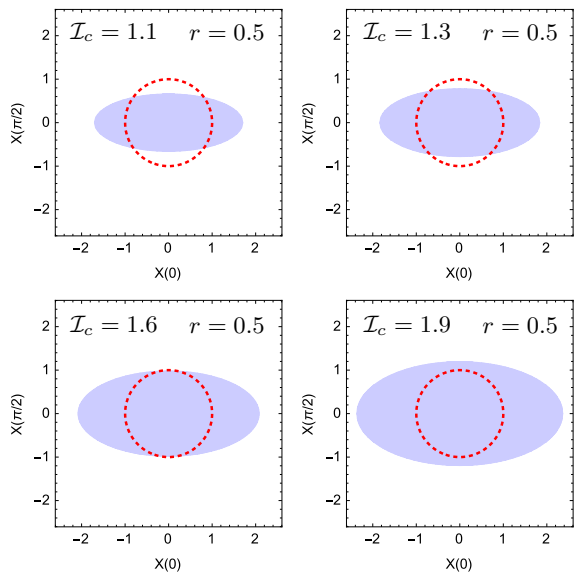
\begin{figure}[ht!]
\begin{tikzpicture}
  \node[anchor=south west,inner sep=0] at (0,0) {\includegraphics[width=7.5 cm]{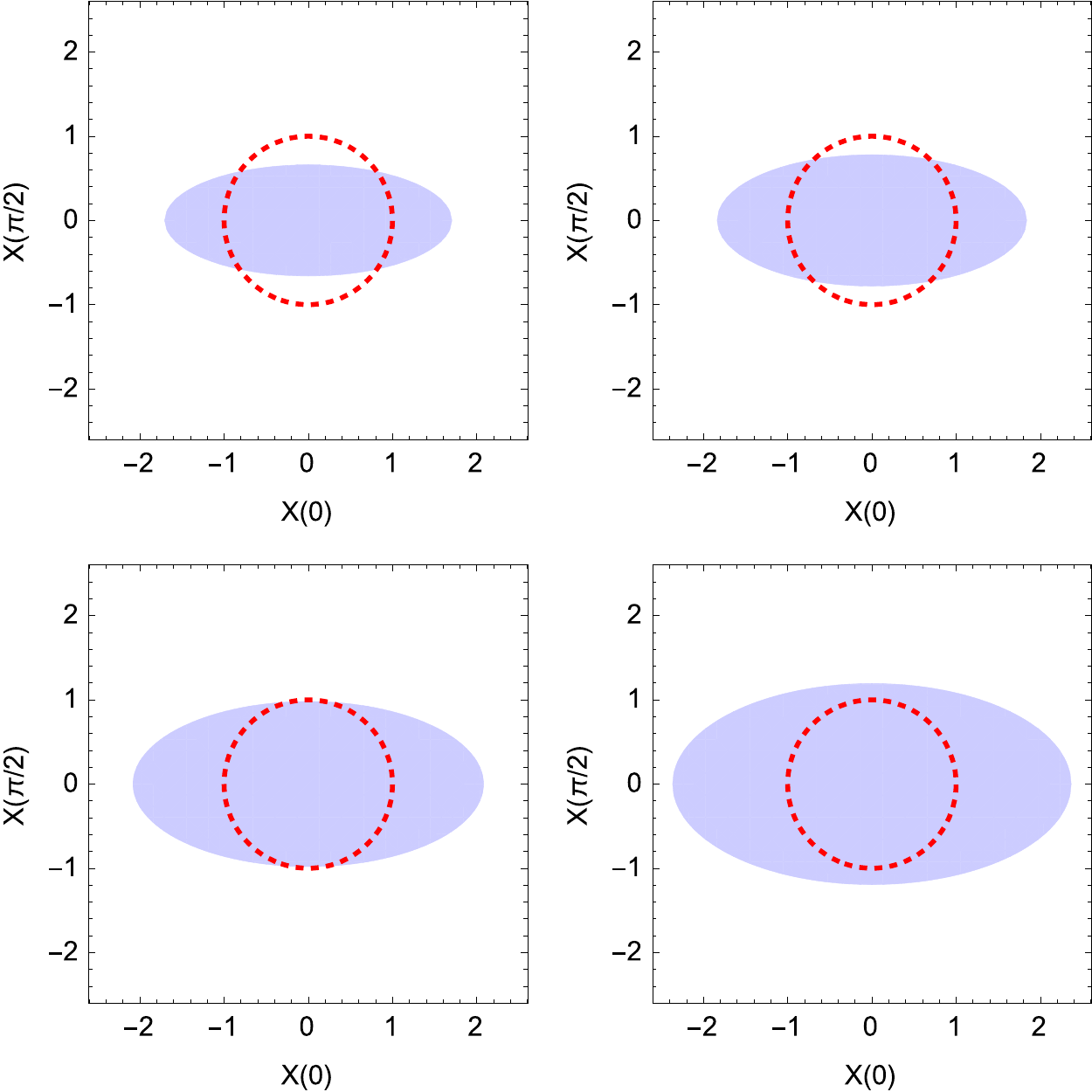}};
  \draw (1.4, 7.2) node {$\mathcal{I}_c = 1.1$};
  \draw (2.9, 7.2) node {$r = 0.5$};
  \draw (5.3, 7.2) node {$\mathcal{I}_c = 1.3$};
  \draw (6.8, 7.2) node {$r = 0.5$};
  
  \draw (1.4, 3.3) node {$\mathcal{I}_c = 1.6$};
  \draw (2.9, 3.3) node {$r = 0.5$};
  \draw (5.3, 3.3) node {$\mathcal{I}_c = 1.9$};
  \draw (6.8, 3.3) node {$r = 0.5$};
\end{tikzpicture}
\caption{\footnotesize Phase space representation of quadrature in the final state. The red dashed circle represents the vacuum 
shot noise, and the blue shaded ellipse represents the quadrature variance of the output state. For fixed single-mode squeezing factor ($r = 0.5$), 
the minimum quadrature variance is below the vacuum shot noise for small $\mathcal{I}_c$, indicating the output state is a squeezed state. While for large
enough $\mathcal{I}_c$, the minimum quadrature variance surpasses the vacuum shot noise, showing that squeezing is destroyed. } 
\label{quadrature}
\end{figure}

\begin{figure}[ht!]
\begin{tikzpicture}
  \node[anchor=south west,inner sep=0] at (0,0) {\includegraphics[width=7.5 cm]{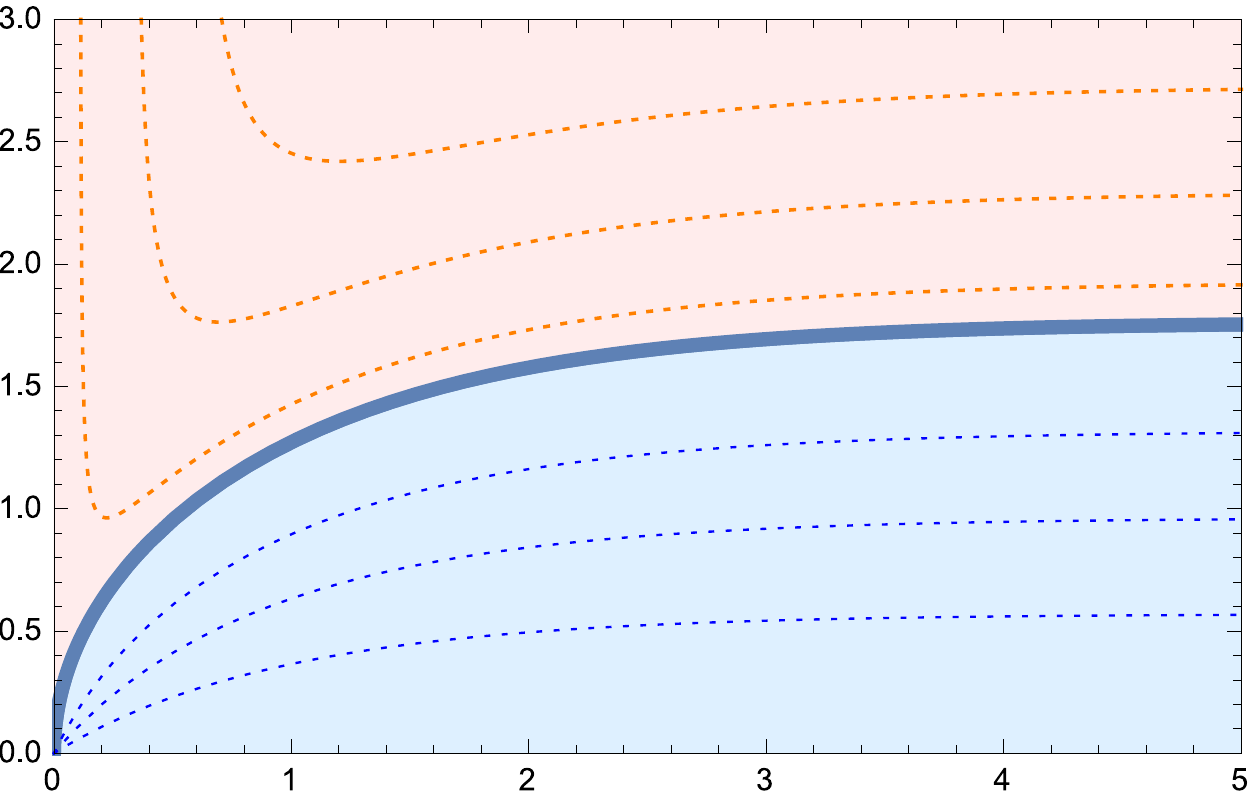}};
  
  \draw  (4, 2.45) node  [rotate=6] {$V_{\text{min}}=1.0$};
  
   \draw  (0.9, 2.2) node {$0.8$};
   \draw  (1.5, 3.2) node {$0.5$};
   \draw  (2, 4.1) node {$0.3$};
   
   \draw (4.5, 3.8) node  {\bf \large \color{orange}Squeezing};
   \draw (3.8, 1.3) node  {\bf \large \color{blue} No Squeezing};

   \draw  (6.5, 2) node {$2.0$};
   \draw  (6.5, 1.5) node {$4.0$};
   \draw  (6.5, 0.9) node {$12.0$};
  
  \draw (4, -0.2) node {$r$};
  \draw (-0.3, 2.5) node [rotate=90] {$2\pi \omega_0/a$};
\end{tikzpicture}
\caption{\footnotesize Distribution of minimum quadrature variance of the output state as a function of single-mode squeezing factor $r$ and the central frequency
$\omega_0$ in the narrow bandwidth limit. A critical curve along which $V_{\text{min}}=1.0$ separates the squeezing region and no squeezing region. 
In the squeezing region  $V_{\text{min}}<1.0$,  while in the no squeezing region  $V_{\text{min}}>1.0$.  } 
\label{MinimumVariance}
\end{figure}
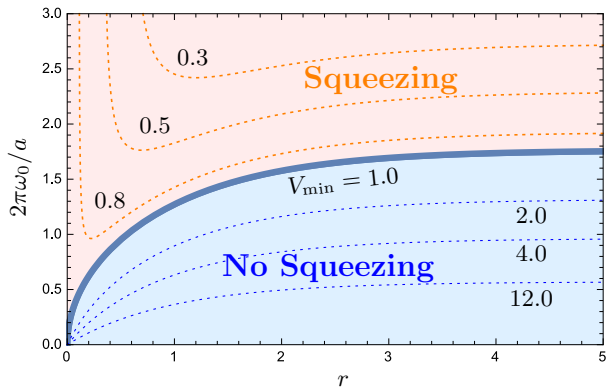

As the Unruh effect in the Rindler frame becomes more pronounced, the decoherence in the Minkowski frame becomes stronger. Eventually squeezing disappears and the final state becomes classical in the sense that coherent state superpositions are removed and the state becomes decomposable into a mixture of coherent states. 
Fig. \ref{quadrature} shows an example of the phase space representation of the quadrature amplitude. 
In the narrow bandwidth limit, we use the approximate relation between $\mathcal{I}_c$ and 
 $\omega_0$ to find the distribution of minimum quadrature variance in terms of $r$ and $\omega_0$, as shown in Fig. \ref{MinimumVariance}. A critical
 curve, which is determined by 
 \begin{equation}\label{criticalCurve}
 \frac{2\pi \omega_0}{a} = \ln \bigg(\frac{\sqrt{1+\coth(r/2)}+1}{\sqrt{1+\coth(r/2)}-1} \bigg),
 \end{equation}
 separates the squeezing region and no squeezing region. 
 When $r \rightarrow \infty$, 
 $2\pi \omega_0/a \rightarrow 2 \ln (\sqrt 2 +1) \approx 1.763$. Below this value, one can always make the output state classical by increasing the 
 single-mode squeezing factor $r$.

 \vspace{0.5cm}
 \noindent \textbf{Entanglement results} 
 
\noindent We generalize the above calculation to a uniformly accelerated two-mode squeezer in the right Rindler wedge that couples the left-moving and right-moving Rindler modes. 
The two-mode squeezing operator is defined as \cite{Scully97}
\begin{eqnarray}\label{two-mode-squeezing}
\hat S_2 (r) = \exp \bigg\{ r \bigg( \hat{b}^{R\dag}_{1g} \hat{b}^{R\dag}_{2g}  - \hat{b}^{R}_{1g} \hat{b}^{R}_{2g} \bigg) \bigg\}, 
\end{eqnarray}
where the subscripts ``1" and ``2" represent the left-moving and right-moving moving modes, respectively. Here $r$ is the squeezing factor and is assumed to be real. The output 
field includes the left-moving and right-moving parts. To have full information about the output state, one needs to measure the states of the left-moving and right-moving modes,
as well as the correlations between them. 

\begin{figure}[ht!]
\begin{tikzpicture}
  \node[anchor=south west,inner sep=0] at (0,0) {\includegraphics[width=7.5 cm]{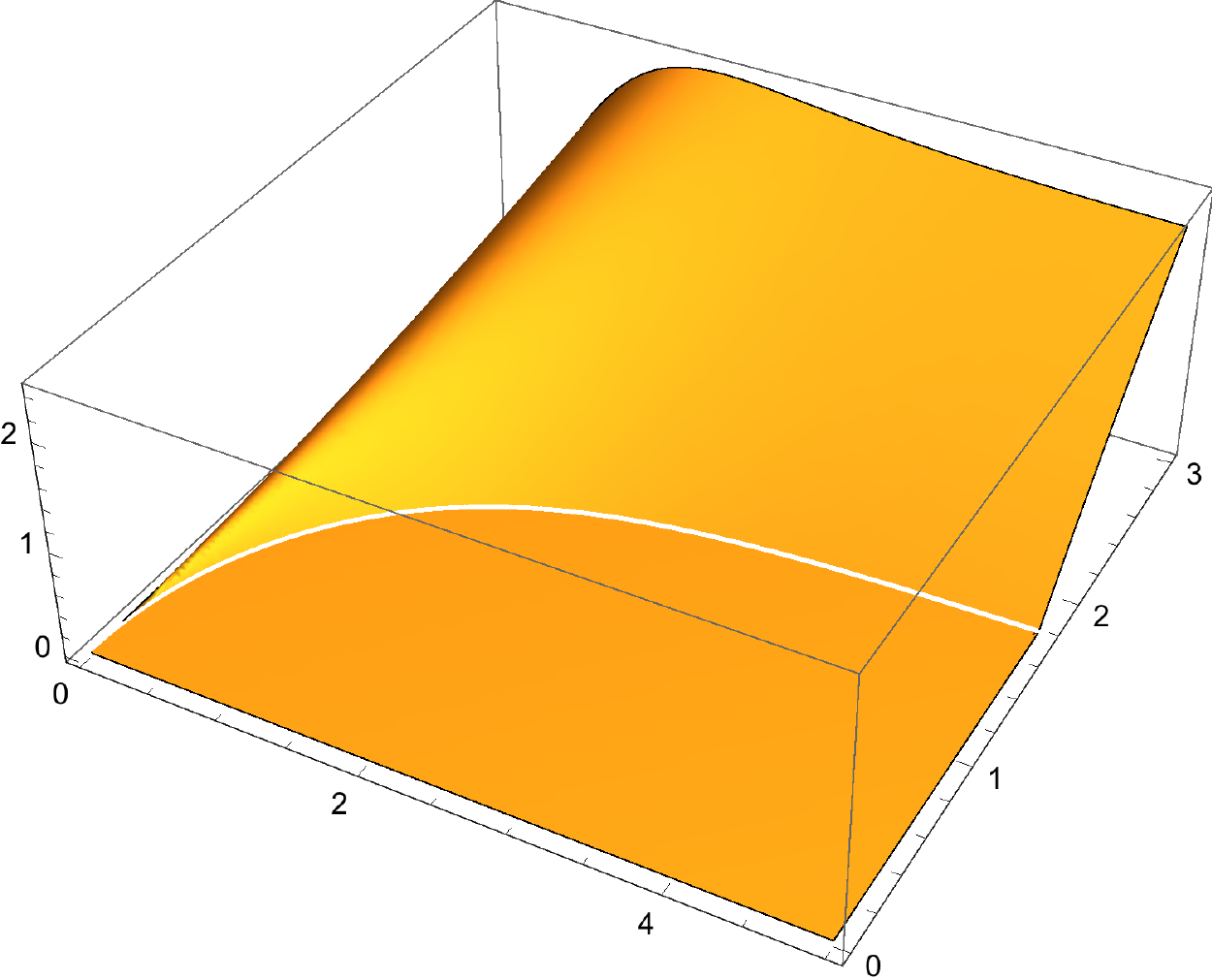}};
  
   \draw (4.5, 4.0) node  {\bf \color{red} Entanglement};
   \draw (3.8, 1.8) node [rotate=-20] {\bf \color{blue!70} No Entanglement};
 
  \draw (-0.2, 2.7) node {$E_{\mathcal{N}}$};
  \draw (2.7, 0.7) node {$r$};
  \draw (6.8, 1.5) node [rotate=60] {$2\pi \omega_0/a$};
  
\end{tikzpicture}
\caption{\footnotesize Logarithmic negativity of the output state as a function of the squeezing factor $r$ and the central frequency
$\omega_0$ in the narrow bandwidth limit.  } 
\label{LogNegativity}
\end{figure}
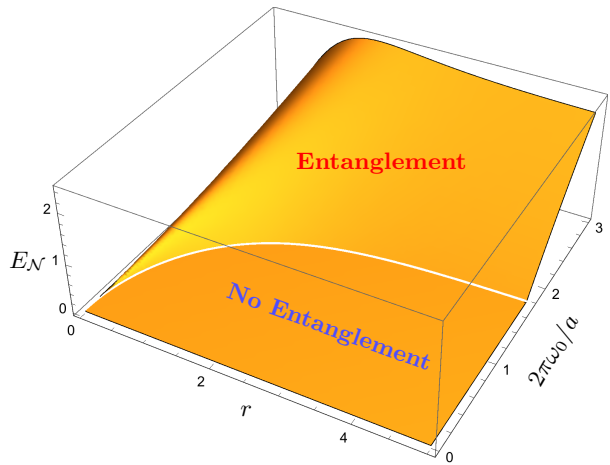

We add two displacements, with amplitudes $\alpha_1 = |\alpha_1|e^{i\phi_1}$ and $\alpha_2 = |\alpha_2|e^{i\phi_2}$, after the two-mode squeezer in order to perform homodyne detection, the former for the left-moving mode and the latter for the right-moving mode. We find that the expectation values of the quadrature amplitudes $\hat X_1(\phi_1)$ and
$\hat X_2(\phi_2)$ are zero. The covariance matrix \cite{Weedbrook12} of the output state is 
\begin{eqnarray}\label{covarMatrix}
\bf V = \begin{pmatrix}
\mathcal{A} & 0 & \mathcal{B} & 0 \\
0 & \mathcal{A} & 0 & -\mathcal{B} \\
\mathcal{B} & 0 & \mathcal{A} & 0 \\
0 & -\mathcal{B} & 0 & \mathcal{A}
\end{pmatrix},
\end{eqnarray}
where 
\begin{eqnarray}
\mathcal{A} &=& (2 \mathcal{I}_c -1)^2 \cosh(2r) - 4\mathcal{I}_c (\mathcal{I}_c - 1)(2 \cosh r - 1), \nonumber\\
\mathcal{B} &=& 2 \sinh r ~\big[ (2 \mathcal{I}_c - 1)^2 \cosh r - 4 \mathcal{I}_c (\mathcal{I}_c - 1) \big].
\end{eqnarray}
From the covariance matrix (\ref{covarMatrix}), one can derive the logarithmic negativity as \cite{Weedbrook12}
\begin{eqnarray}
E_{\mathcal{N}} = \text{max}[0, -\log_2(\tilde \nu_-)],
\end{eqnarray}
where $\tilde \nu_-$ is the smallest symplectic eigenvalue of the partially transposed state,
\begin{eqnarray}
\tilde \nu_- = e^{-2r} + 4\mathcal{I}_c (\mathcal{I}_c-1) (e^{-r}-1)^2. 
\label{ee}
\end{eqnarray}
When $\tilde \nu_- < 1$ ($E_{\mathcal{N}} > 0$), there exists entanglement between the left-moving and right-moving modes; when $\tilde \nu_- \ge 1$ ($E_{\mathcal{N}} = 0$),
the left-moving and right-moving modes are not entangled. 
When $\mathcal{I}_c = 1$ the covariance matrix (equation \ref{covarMatrix}) is that of a pure two-mode squeezed state and the entanglement (equation \ref{ee}) is maximised. However, when $\mathcal{I}_c > 1$ the covariance matrix becomes decohered (mixed) and the entanglement drops, eventually disappearing.
Fig. \ref{LogNegativity} shows the logarithmic negativity as a function of the squeezing factor $r$ and the 
central frequency $\omega_0$ in the narrow bandwidth limit. The critical curve $\tilde \nu_- = 1$, dividing the entanglement and no entanglement regions, is determined by 
equation (\ref{criticalCurve}).

 \vspace{0.5cm}
 \noindent \textbf{Conclusion} 
 
\noindent The decoherence effect we describe here is a previously unnoticed consequence of the transformation from the bipartite Hilbert space of the Rindler and Unruh modes, to the single Hilbert space of the Minkowski modes. Notice that in equation (\ref{ParcleNumber}) any direct phase relationship between the left and right Unruh modes is lost in the construction of the Minkowski number operator. This means that interactions which lead to entanglement between the left and right Unruh modes, as occurs with the accelerated squeezer and the entangler, will in general appear as decoherence with respect to measurements by inertial observers. In contrast, coherent state signals do not produce Unruh mode entanglement and so no decoherence is observed for such signals.

We have shown that single and two-mode mode unitary squeezing operations in an accelerated frame are in general detected as decohered states by inertial observers. As we noted in the introduction, the standard Unruh effect can be purified if a mirror image accelerated observer is introduced. Here we find that a mirror image accelerated source is required to purify the state detected by the inertial observer. In particular, for the narrow band case, only if the mirror image source displaces the state by $\gamma = 2{{\sqrt{\mathcal{I}_c(\mathcal{I}_c - 1)}}\over{2\mathcal{I}_c-1}} \alpha^*$, in phase with the original accelerated source, then the inertial detectors will see pure states in both the squeezer and entangler cases. Details of this calculation are given in the appendix.
 
We believe the decoherence effect has significance for understanding quantum effects in gravitational systems. For example, our system can be viewed as a toy model for the creation and eventual evaporation of a black-hole. We begin in the far past in a pure Minkowski vacuum state, before the formation of the black-hole. In the intermediate epoch accelerated observers, representing observers close to the black-hole, interact with the field modes.
Finally in the far-future the black-hole has evaporated leaving flat space, however the field is left in a mixed state with respect to inertial observers. This may indicate a new direction for understanding the black-hole information paradox. 
 
 The accelerations required to generate this decoherence effect are well beyond those that can be physically produced in the lab. However, such accelerations do occur naturally in many regions of the universe. In addition the equivalence between acceleration and time dependent effects \cite{OLS11} may enable laboratory tests, especially at micro-wave frequencies \cite{WIL11}. We also note that simulation of these effects using optical squeezing is possible with current technology and would allow an experimental investigation of analogues to the decoherence effect described here that may be of interest in their own right from a quantum information point of view.

%
\vspace{0.5cm}
\section*{Appendix}
\subsection{Derivation of the variance in single-mode squeezing case}

In the single-mode squeezing case, the unitary is taken to be $\hat{U}_g = \hat{D}_g(\alpha) \hat{S}_{1}(r)$.
By substituting this unitary into equation (\ref{GeneralUnruh}) one can derive the input-output relations for Unruh modes,
\begin{eqnarray}\label{IOUnruh}
\hat{c}^{\prime}_{\omega} &=& \hat{c}_{\omega} + g^*(\omega) \cosh r_{\omega} \big[ \hat{b}^R_g(\cosh r -1)  + \hat{b}^{R\dag}_g \sinh r + \alpha \big], \nonumber\\
\hat{d}^{\prime}_{\omega} &=& \hat{d}_{\omega} - g(\omega) \sinh r_{\omega} \big[ \hat{b}^{R \dag}_g(\cosh r -1)  + \hat{b}^{R}_g \sinh r + \alpha^* \big].
\end{eqnarray}
The localized Rindler operator $\hat{b}^R_g$ can be expressed in terms of the input Unruh operators by using the transformations between the Rindler and 
Unruh modes. Equation (\ref{IOUnruh}) becomes
\begin{eqnarray}
\hat{c}^{\prime}_{\omega} &=& \hat{c}_{\omega} + g^*(\omega) \cosh r_{\omega} \bigg[ 
(\cosh r -1) \int \mathrm{d} \omega' g(\omega') \big( \hat{c}_{\omega'} \cosh r_{\omega'} + \hat{d}^{\dag}_{\omega'} \sinh r_{\omega'} \big) \nonumber \\
&&+  \sinh r \int \mathrm{d} \omega' g^*(\omega') \big( \hat{c}^{\dag}_{\omega'} \cosh r_{\omega'} + \hat{d}_{\omega'} \sinh r_{\omega'} \big) + \alpha 
\bigg], \nonumber\\
\nonumber\\
\hat{d}^{\prime}_{\omega} &=& \hat{d}_{\omega} - g(\omega) \sinh r_{\omega} \bigg[ 
(\cosh r -1) \int \mathrm{d} \omega' g^*(\omega') \big( \hat{c}^{\dag}_{\omega'} \cosh r_{\omega'} + \hat{d}_{\omega'} \sinh r_{\omega'} \big) \nonumber\\
&&+ \sinh r \int \mathrm{d} \omega' g(\omega') \big( \hat{c}_{\omega'} \cosh r_{\omega'} + \hat{d}^{\dag}_{\omega'} \sinh r_{\omega'} \big) + \alpha^* 
\bigg].
\end{eqnarray}
It is now straightforward to calculate the vacuum expectation values of the product of two
output Unruh operators.
\begin{eqnarray}
\langle 0 | \hat{c}^{\prime \dag}_{\omega} \hat{c}^{\prime}_{\omega'} | 0 \rangle &=& g(\omega) g^*(\omega') \cosh r_{\omega} \cosh r_{\omega'} (E_c+|\alpha|^2), \nonumber \\
\langle 0 | \hat{d}^{\prime \dag}_{\omega} \hat{d}^{\prime}_{\omega'} | 0 \rangle &=& g^*(\omega) g(\omega') \sinh r_{\omega} \sinh r_{\omega'} (E_d+|\alpha|^2), \nonumber \\
\langle 0 | \hat{c}^{\prime}_{\omega} \hat{c}^{\prime}_{\omega'} | 0 \rangle &=& g^*(\omega) g^*(\omega') \cosh r_{\omega} \cosh r_{\omega'} (E_{cc}+\alpha^2), \nonumber\\
\langle 0 | \hat{d}^{\prime}_{\omega} \hat{d}^{\prime}_{\omega'} | 0 \rangle &=& g(\omega) g(\omega') \sinh r_{\omega} \sinh r_{\omega'}(E_{dd}+\alpha^{*2}), \nonumber\\
\langle 0 | \hat{c}^{\prime}_{\omega} \hat{d}^{\prime}_{\omega'} | 0 \rangle &=& g^*(\omega) g(\omega') \cosh r_{\omega} \sinh r_{\omega'} (E_{cd}-|\alpha|^2), \nonumber\\
\langle 0 | \hat{c}^{\prime \dag}_{\omega} \hat{d}^{\prime}_{\omega'} | 0 \rangle &=& g(\omega) g(\omega') \cosh r_{\omega}\sinh r_{\omega'} (\bar{E}_{cd}-\alpha^{*2}),
\end{eqnarray}
where 
\begin{eqnarray}\label{E:quantities}
E_c &=& \mathcal{I}_s (\cosh r -1)^2 + \mathcal{I}_c \sinh^2 r, \nonumber\\
E_d &=&  \mathcal{I}_c (\cosh r -1)^2 + \mathcal{I}_s \sinh^2 r, \nonumber\\
E_{cc} &=& \sinh r \big[ (\mathcal{I}_c + \mathcal{I}_s) (\cosh r -1) + 1 \big], \nonumber\\
E_{dd} &=& \sinh r \big[ (\mathcal{I}_c + \mathcal{I}_s) (\cosh r -1) - 1 \big], \nonumber\\
E_{cd} &=& - \cosh r  (\cosh r -1) (\mathcal{I}_c + \mathcal{I}_s), \nonumber\\
\bar{E}_{cd} &=& - \sinh r  (\cosh r -1) (\mathcal{I}_c + \mathcal{I}_s).
\end{eqnarray}
Other vacuum expectation values are either zero or complex conjugates of the above ones.
From equations (\ref{ParcleNumber}) and (\ref{SquareNumber}), the vacuum expectation value of the total Minkowski particle number is 
\begin{eqnarray}\label{totalparticle}
 \langle 0 | \hat{N} | 0 \rangle 
&=& |\alpha|^2 (\mathcal{I}_c + \mathcal{I}_s)  + (\mathcal{I}_c E_c + \mathcal{I}_s E_d)
\end{eqnarray}
and the variance of total Minkowski particle number is
\begin{eqnarray}
 (\Delta N)^2 &=& \langle 0 | \hat{N}^2 | 0 \rangle -  \langle 0 | \hat{N} | 0 \rangle^2 \nonumber\\
&=&  |\alpha|^2 \bigg[( \mathcal{I}_c + \mathcal{I}_s) + 2 ( \mathcal{I}_c^2 E_c + \mathcal{I}_s^2 E_d ) + 2\mathcal{I}_c^2 E_{cc} \cos (2\phi) \nonumber\\
&&+ 2\mathcal{I}_s^2 E_{dd} \cos (2\phi) - 4 \mathcal{I}_c \mathcal{I}_s E_{cd} - 4 \mathcal{I}_c \mathcal{I}_s \bar{E}_{cd} \cos (2\phi)\bigg],
\label{Nsquared}
\end{eqnarray}
where $\phi$ is the displacement phase. In the homodyne detection, normalizing the 
variance of the particle number using the strength of the local oscillator gives the variance of the quadrature amplitude \cite{BachorRalph}. Here the 
strength of the local oscillator is $\sim |\alpha|^2(\mathcal{I}_c + \mathcal{I}_s)$, so the variance of quadrature amplitude is 
\begin{eqnarray}
V(\phi) &=& \frac{ (\Delta N)^2 }{ |\alpha|^2(\mathcal{I}_c + \mathcal{I}_s) } \nonumber\\
&=& \cosh(2r) + 4 \mathcal{I}_c (\mathcal{I}_c-1) (\cosh 2r -2 \cosh r + 1) 
+ 2 \sinh r \big[(2\mathcal{I}_c -1)^2 \cosh r - 4 \mathcal{I}_c (\mathcal{I}_c-1)\big] \cos(2\phi). 
\end{eqnarray}

\subsection{Uniformly accelerated two-mode squeezer}

For a massless scalar field, the left-moving and right-moving Rindler modes are decoupled.
We consider a uniformly accelerated two-mode squeezer in the right Rindler wedge that couples the left-moving and right-moving Rindler modes. 
Entanglement between the left-moving and right-moving Rindler modes might be created by the accelerated two-mode squeezer. One question of particular interest is, 
given that entanglement has been created as viewed
by uniformly accelerated observers, whether entanglement between left-moving and right-moving fields exists as observed by inertial observers. 

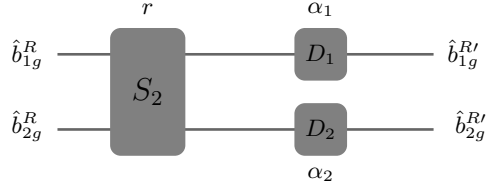
\begin{figure}[ht!]

\centering
\begin{tikzpicture}[scale=1.]

\begin{scope}[line width = 1pt, color=black!60] 
\draw (0.5, 0.3) -- (5.5, 0.3);
\draw (0.5, 1.3) -- (5.5, 1.3);
\end{scope}

\begin{scope} [color = gray, rounded corners, rotate=0]
\fill (1.2, -0.05) rectangle +(1, 1.7);
\fill (3.65, 0.95) rectangle +(0.7, 0.7);
\fill (3.65, -0.05) rectangle +(0.7, 0.7);
\end{scope}

\begin{scope}
\draw (0.1, 0.35) node {$\hat b^R_{2g}$};
\draw (0.1, 1.3) node {$\hat b^R_{1g}$};

\draw (6, 0.35) node {$\hat b^{R \prime}_{2g}$};
\draw (5.9, 1.3) node {$\hat b^{R \prime}_{1g}$};
\end{scope}

\begin{scope}
\draw (1.7, 0.8) node {\large $S_{2}$};
\draw (4, 1.3) node { $D_1$};
\draw (4, 0.3) node { $D_2$};
\draw (4, 1.9) node { $\alpha_1$};
\draw (4, -0.3) node { $\alpha_2$};
\draw (1.7, 1.9) node { $r$};
\end{scope}

\end{tikzpicture}
\caption{\footnotesize A uniformly accelerated two-mode squeezer and the scheme of self-homodyne detection.  } 
\label{two-mode}
\end{figure}

The unitary characterizing the uniformly accelerated two-mode squeezer is given by equation (\ref{two-mode-squeezing}). Similar to the case of the uniformly accelerated 
single-mode squeezer, two uniformly accelerated displacements, $\hat D_1(\alpha_1)$ and $\hat D_2(\alpha_2)$, are introduced to realize the homodyne detection,
where $\alpha_1 = |\alpha_1| e^{i \phi_1}$ and $\alpha_2 = |\alpha_2| e^{i \phi_2}$. 
The relevant circuit is shown in Fig. \ref{two-mode}. The two local oscillators are used to detect the left-moving and right-moving fields, as well as the correlations between them. 
In this appendix, we are going to derive the covariance matrix for the output state of the uniformly accelerated two-mode squeezer. 

The input-output relations for the circuit Fig. \ref{two-mode} can be derived straightforwardly. By further using the relations between the Rindler operators and Unruh operators \cite{Su16}, we can derive the input-output relations for the Unruh operators.
\begin{eqnarray}
\hat{c}^{\prime}_{1 \omega} &=& \hat{c}_{1 \omega} + g^*(\omega) \cosh r_{\omega} \bigg[ 
(\cosh r -1) \int \mathrm{d} \omega' g(\omega') \big( \hat{c}_{1 \omega'} \cosh r_{\omega'} + \hat{d}^{\dag}_{1\omega'} \sinh r_{\omega'} \big) \nonumber \\
&&+  \sinh r \int \mathrm{d} \omega' g^*(\omega') \big( \hat{c}^{\dag}_{2 \omega'} \cosh r_{\omega'} + \hat{d}_{2 \omega'} \sinh r_{\omega'} \big) + \alpha_1 
\bigg], \nonumber\\
\nonumber\\
\hat{d}^{\prime}_{1 \omega} &=& \hat{d}_{1 \omega} - g(\omega) \sinh r_{\omega} \bigg[ 
(\cosh r -1) \int \mathrm{d} \omega' g^*(\omega') \big( \hat{c}^{\dag}_{1 \omega'} \cosh r_{\omega'} + \hat{d}_{1 \omega'} \sinh r_{\omega'} \big) \nonumber\\
&&+ \sinh r \int \mathrm{d} \omega' g(\omega') \big( \hat{c}_{2 \omega'} \cosh r_{\omega'} + \hat{d}^{\dag}_{2 \omega'} \sinh r_{\omega'} \big) + \alpha^*_1
\bigg]. \nonumber\\
\nonumber\\
\hat{c}^{\prime}_{2 \omega} &=& \hat{c}_{2 \omega} + g^*(\omega) \cosh r_{\omega} \bigg[ 
(\cosh r -1) \int \mathrm{d} \omega' g(\omega') \big( \hat{c}_{2 \omega'} \cosh r_{\omega'} + \hat{d}^{\dag}_{2 \omega'} \sinh r_{\omega'} \big) \nonumber \\
&&+  \sinh r \int \mathrm{d} \omega' g^*(\omega') \big( \hat{c}^{\dag}_{1 \omega'} \cosh r_{\omega'} + \hat{d}_{1 \omega'} \sinh r_{\omega'} \big) + \alpha_2 
\bigg], \nonumber\\
\nonumber\\
\hat{d}^{\prime}_{2 \omega} &=& \hat{d}_{2 \omega} - g(\omega) \sinh r_{\omega} \bigg[ 
(\cosh r -1) \int \mathrm{d} \omega' g^*(\omega') \big( \hat{c}^{\dag}_{2 \omega'} \cosh r_{\omega'} + \hat{d}_{2 \omega'} \sinh r_{\omega'} \big) \nonumber\\
&&+ \sinh r \int \mathrm{d} \omega' g(\omega') \big( \hat{c}_{1 \omega'} \cosh r_{\omega'} + \hat{d}^{\dag}_{1 \omega'} \sinh r_{\omega'} \big) + \alpha^*_2
\bigg].
\end{eqnarray}
It is then straightforward to calculate the vacuum expectation values of the products of two output Unruh operators. For the left-moving operators, we have
\begin{eqnarray}\label{EVleft}
\langle 0 | \hat{c}^{\prime \dag}_{1 \omega} \hat{c}^{\prime}_{1 \omega'} | 0 \rangle &=& g(\omega) g^*(\omega') \cosh r_{\omega} \cosh r_{\omega'} (E_c+|\alpha_1|^2), \nonumber \\
\langle 0 | \hat{d}^{\prime \dag}_{1 \omega} \hat{d}^{\prime}_{1 \omega'} | 0 \rangle &=& g^*(\omega) g(\omega') \sinh r_{\omega} \sinh r_{\omega'} (E_d+|\alpha_1|^2), \nonumber \\
\langle 0 | \hat{c}^{\prime}_{1 \omega} \hat{c}^{\prime}_{1 \omega'} | 0 \rangle &=& \alpha_1^2 ~ g^*(\omega) g^*(\omega') \cosh r_{\omega} \cosh r_{\omega'}, \nonumber\\
\langle 0 | \hat{d}^{\prime}_{1 \omega} \hat{d}^{\prime}_{1 \omega'} | 0 \rangle &=& \alpha_1^{*2} ~ g(\omega) g(\omega') \sinh r_{\omega} \sinh r_{\omega'}, \nonumber\\
\langle 0 | \hat{c}^{\prime}_{1 \omega} \hat{d}^{\prime}_{1 \omega'} | 0 \rangle &=& g^*(\omega) g(\omega') \cosh r_{\omega} \sinh r_{\omega'} (E_{cd}-|\alpha_1|^2), \nonumber\\
\langle 0 | \hat{c}^{\prime \dag}_{1 \omega} \hat{d}^{\prime}_{1 \omega'} | 0 \rangle &=& -\alpha_1^{*2} ~ g(\omega) g(\omega') \cosh r_{\omega}\sinh r_{\omega'}.
\end{eqnarray}
For the right-moving operators, we have
\begin{eqnarray}
\langle 0 | \hat{c}^{\prime \dag}_{2 \omega} \hat{c}^{\prime}_{2 \omega'} | 0 \rangle &=& g(\omega) g^*(\omega') \cosh r_{\omega} \cosh r_{\omega'} (E_c+|\alpha_2|^2), \nonumber \\
\langle 0 | \hat{d}^{\prime \dag}_{2 \omega} \hat{d}^{\prime}_{2 \omega'} | 0 \rangle &=& g^*(\omega) g(\omega') \sinh r_{\omega} \sinh r_{\omega'} (E_d+|\alpha_2|^2), \nonumber \\
\langle 0 | \hat{c}^{\prime}_{2 \omega} \hat{c}^{\prime}_{2 \omega'} | 0 \rangle &=& \alpha_2^2 ~ g^*(\omega) g^*(\omega') \cosh r_{\omega} \cosh r_{\omega'}, \nonumber\\
\langle 0 | \hat{d}^{\prime}_{2 \omega} \hat{d}^{\prime}_{2 \omega'} | 0 \rangle &=& \alpha_2^{*2} ~ g(\omega) g(\omega') \sinh r_{\omega} \sinh r_{\omega'}, \nonumber\\
\langle 0 | \hat{c}^{\prime}_{2 \omega} \hat{d}^{\prime}_{2 \omega'} | 0 \rangle &=& g^*(\omega) g(\omega') \cosh r_{\omega} \sinh r_{\omega'} (E_{cd}-|\alpha_2|^2), \nonumber\\
\langle 0 | \hat{c}^{\prime \dag}_{2 \omega} \hat{d}^{\prime}_{2 \omega'} | 0 \rangle &=& -\alpha_2^{*2} ~ g(\omega) g(\omega') \cosh r_{\omega}\sinh r_{\omega'}.
\end{eqnarray}
For the products of the left-moving and right-moving operators, we have
\begin{eqnarray}
\langle 0 | \hat{c}^{\prime \dag}_{1 \omega} \hat{c}^{\prime}_{2 \omega'} | 0 \rangle &=& \alpha_1^* \alpha_2 ~ g(\omega) g^*(\omega') \cosh r_{\omega} \cosh r_{\omega'}, \nonumber \\
\langle 0 | \hat{d}^{\prime \dag}_{1 \omega} \hat{d}^{\prime}_{2 \omega'} | 0 \rangle &=& \alpha_1 \alpha_2^* ~ g^*(\omega) g(\omega') \sinh r_{\omega} \sinh r_{\omega'}, \nonumber \\
\langle 0 | \hat{c}^{\prime}_{1 \omega} \hat{c}^{\prime}_{2 \omega'} | 0 \rangle &=& g^*(\omega) g^*(\omega') \cosh r_{\omega} \cosh r_{\omega'}(E_{cc} +  \alpha_1 \alpha_2 ), \nonumber\\
\langle 0 | \hat{d}^{\prime}_{1 \omega} \hat{d}^{\prime}_{2 \omega'} | 0 \rangle &=& g(\omega) g(\omega') \sinh r_{\omega} \sinh r_{\omega'}(E_{dd} +  \alpha_1^* \alpha_2^* ), \nonumber\\
\langle 0 | \hat{c}^{\prime}_{1 \omega} \hat{d}^{\prime}_{2 \omega'} | 0 \rangle &=&- \alpha_1 \alpha_2^* ~  g^*(\omega) g(\omega') \cosh r_{\omega} \sinh r_{\omega'}, \nonumber\\
\langle 0 | \hat{d}^{\prime}_{1 \omega} \hat{c}^{\prime}_{2 \omega'} | 0 \rangle &=&- \alpha_1^* \alpha_2 ~  g(\omega) g^*(\omega') \sinh r_{\omega} \cosh r_{\omega'}, \nonumber\\
\langle 0 | \hat{c}^{\prime \dag}_{1 \omega} \hat{d}^{\prime}_{2 \omega'} | 0 \rangle &=& g(\omega) g(\omega') \cosh r_{\omega}\sinh r_{\omega'} (\bar E_{cd} - \alpha_1^* \alpha_2^*). \nonumber\\
\langle 0 | \hat{d}^{\prime \dag}_{1 \omega} \hat{c}^{\prime}_{2 \omega'} | 0 \rangle &=& g^*(\omega) g^*(\omega') \sinh r_{\omega}\cosh r_{\omega'} (\bar E_{cd} - \alpha_1 \alpha_2).
\end{eqnarray}
The vacuum expectation values of the left-moving and right-moving Minkowski particle number are
\begin{eqnarray}\label{totalparticle}
\langle 0 | \hat{N}_1 | 0 \rangle &=& |\alpha_1|^2 (\mathcal{I}_c + \mathcal{I}_s)  + (\mathcal{I}_c E_c + \mathcal{I}_s E_d), \nonumber\\
\langle 0 | \hat{N}_2 | 0 \rangle &=& |\alpha_2|^2 (\mathcal{I}_c + \mathcal{I}_s)  + (\mathcal{I}_c E_c + \mathcal{I}_s E_d).
\end{eqnarray}
To accomplish the homodyne detection, we also need to know the strength of the local oscillators in the absence of signal: 
$\langle 0 | \hat{N}_{10} | 0 \rangle = |\alpha_1|^2 (\mathcal{I}_c + \mathcal{I}_s)$, $\langle 0 | \hat{N}_{20} | 0 \rangle = |\alpha_2|^2 (\mathcal{I}_c + \mathcal{I}_s)$.
According to equation (\ref{quad}), the expectation values of the left-moving and right-moving quadrature amplitudes can be found as
\begin{eqnarray}
\langle 0 | \hat X_1(\phi_1) | 0 \rangle = \langle 0 | \hat X_2(\phi_2) | 0 \rangle = 0
\end{eqnarray}
in the limit of $|\alpha_1| \gg 1$ and $|\alpha_2| \gg 1$.

Using equation (\ref{EVleft}) and keeping terms to second order of $\alpha_1$, we find 
\begin{eqnarray}
&&\langle 0 | \hat N_1(\phi_1)  \hat N_1(\phi_1^{\prime}) | 0 \rangle - \langle 0 | \hat N_1(\phi_1) | 0 \rangle \langle 0 | \hat N_1(\phi_1^{\prime}) | 0 \rangle \nonumber\\
&=& \alpha_1^*\alpha_1^{\prime} \mathcal{I}_c + \alpha_1 \alpha_1^{\prime *} \mathcal{I}_s + (\alpha_1^*\alpha_1^{\prime} + \alpha_1 \alpha_1^{\prime *})
(\mathcal{I}_c^2 E_c + \mathcal{I}_s^2 E_d - 2 \mathcal{I}_c \mathcal{I}_s E_{cd}).
\end{eqnarray}
When $\phi_1 = \phi_1^{\prime}$, we obtain the variance of the left-moving quadrature amplitude,
\begin{eqnarray}\label{VarianceLeft}
V_1(\phi_1) = \frac{\langle 0 | \hat N^2_1(\phi_1) | 0 \rangle - \langle 0 | \hat N_1(\phi_1) | 0 \rangle^2} {\langle 0 | \hat N_{10}(\phi_1) | 0 \rangle}
= 1 + 2 (\mathcal{I}_c^2 E_c + \mathcal{I}_s^2 E_d - 2 \mathcal{I}_c \mathcal{I}_s E_{cd}) / (\mathcal{I}_c + \mathcal{I}_s). 
\end{eqnarray}
We are also interested in the case where $\phi_1^{\prime} = \phi_1 + \pi/2$. 
\begin{eqnarray}
\langle 0 | \hat X_1(\phi_1)  \hat X_1(\phi_1 + \pi/2) | 0 \rangle 
&=&\frac{\langle 0 | \hat N_1(\phi_1)  \hat N_1(\phi_1+\pi/2) | 0 \rangle - \langle 0 | \hat N_1(\phi_1) | 0 \rangle \langle 0 | \hat N_1(\phi_1+\pi/2) | 0 \rangle}
{\sqrt{\langle 0 | \hat N_{10}(\phi_1) | 0 \rangle \langle 0 | \hat N_{10}(\phi_1+\pi/2) | 0 \rangle}} \nonumber\\
&=& \frac{i}{\mathcal{I}_c + \mathcal{I}_s}.
\end{eqnarray}
According to the symmetry between the left-moving and right-moving modes, similar results can be obtained:
\begin{eqnarray}
V_2(\phi_2) &=& 1 + 2 (\mathcal{I}_c^2 E_c + \mathcal{I}_s^2 E_d - 2 \mathcal{I}_c \mathcal{I}_s E_{cd}) / (\mathcal{I}_c + \mathcal{I}_s)
\end{eqnarray}
and
\begin{eqnarray}
\langle 0 | \hat X_2(\phi_2)  \hat X_2(\phi_2 + \pi/2) | 0 \rangle &=&  \frac{i}{\mathcal{I}_c + \mathcal{I}_s}.
\end{eqnarray}
To first order of $\alpha_1 \alpha_2$, we find
\begin{eqnarray}
\langle 0 | \hat N_1(\phi_1)  \hat N_2(\phi_2) | 0 \rangle - \langle 0 | \hat N_1(\phi_1) | 0 \rangle \langle 0 | \hat N_2(\phi_2) | 0 \rangle
&=& (\alpha_1 \alpha_2 + \alpha_1^* \alpha_2^*) (\mathcal{I}_c^2 E_{cc} + \mathcal{I}_s^2 E_{dd} - 2 \mathcal{I}_c \mathcal{I}_s \bar E_{cd}). 
\end{eqnarray}
The vacuum expectation value of the product of the left-moving and right-moving quadrature amplitudes is 
\begin{eqnarray}\label{Correlations}
\langle 0 | \hat X_1(\phi_1)  \hat X_2(\phi_2) | 0 \rangle 
&=& \frac{\langle 0 | \hat N_1(\phi_1)  \hat N_2(\phi_2) | 0 \rangle - \langle 0 | \hat N_1(\phi_1) | 0 \rangle \langle 0 | \hat N_2(\phi_2) | 0 \rangle}
{\sqrt{\langle 0 | \hat N_{10}(\phi_1) | 0 \rangle \langle 0 | \hat N_{20}(\phi_2) | 0 \rangle}} \nonumber\\
&=& 2 (\mathcal{I}_c^2 E_{cc} + \mathcal{I}_s^2 E_{dd} - 2 \mathcal{I}_c \mathcal{I}_s \bar E_{cd}) \cos (\phi_1+\phi_2)/(\mathcal{I}_c + \mathcal{I}_s). 
\end{eqnarray}

For Gaussian states, the covariance matrix is a very important quantity to characterize the state. In the special case where the expectation values of the quadrature 
amplitudes are zero, which is case that we are considering, the covariance matrix contains full information of the state. We formally define an operator vector
\begin{eqnarray}
\hat {\bf x} \equiv (\hat x_1, \hat p_1, \hat x_2, \hat p_2)^{\text{T}} = \big(\hat X_1(0), \hat X_1(\pi/2), \hat X_2(0), \hat X_2(\pi/2) \big)^{\text{T}}. 
\end{eqnarray}
The covariance matrix $\bf V$ is defined as
\begin{eqnarray}
V_{ij} = \frac{1}{2} \langle \{ \delta \hat x_i, \delta \hat x_j \}\rangle,
\end{eqnarray}
where $\delta \hat x_i = \hat x_i - \langle \hat x_i \rangle $ and $\{~, ~\}$ is the anticommutator. Using the fact that $\langle \hat x_i \rangle = 0$ and equations 
(\ref{VarianceLeft})-(\ref{Correlations}), we find the nonvanishing components of the covariance matrix are
\begin{eqnarray}
V_{11} &=& V_{22} = V_{33} = V_{44} 
= 1 + 2 (\mathcal{I}_c^2 E_c + \mathcal{I}_s^2 E_d - 2 \mathcal{I}_c \mathcal{I}_s E_{cd}) / (\mathcal{I}_c + \mathcal{I}_s). \nonumber\\
V_{13} &=& V_{31} = -V_{24} = -V_{42} 
=  2 (\mathcal{I}_c^2 E_{cc} + \mathcal{I}_s^2 E_{dd} - 2 \mathcal{I}_c \mathcal{I}_s \bar E_{cd})/(\mathcal{I}_c + \mathcal{I}_s).
\end{eqnarray}
By using equation (\ref{E:quantities}) and the relation $\mathcal{I}_c - \mathcal{I}_s = 1$, we obtain the covariance matrix equation (\ref{covarMatrix}).

\vspace{0.5cm}
\subsection{An additional displacement in the left Rindler wedge}

Whilst, in general it is not possible for the inertial detector to see a pure state if the squeezer and local oscillator are imposed in the right Rindler wedge, in this appendix, we show that by appropriately adding an additional local oscillator in the left Rindler wedge, the inertial detector can see a pure state. 
Physically this would require a mirror-image accelerated source to perform coordinated displacements of the quantum field in their reference frame.

In order to match the mode shape in the right Rindler wedge, the wave packet mode in the left Rindler wedge is chosen as $g^*(\omega)$.
The appearance of the complex conjugate comes from the fact that the coordinate time in the left Rindler wedge runs backward compared to that in the right Rindler wedge (and the Minkowski time coordinate). The displacement operator is thus
$\hat D_{g^*}(\gamma)$, where $\gamma = |\gamma| e^{i\phi_{\gamma}}$ and $\phi_{\gamma}$ is the phase. We further require that the phase $\phi_{\gamma}$
satisfies $\phi_{\gamma} = - \phi$. For convenience, we define the ratio between the amplitude of the displacements in the left and right Rindler wedges as
$z \equiv |\gamma|/|\alpha|$. From the general input-output relations of the Unruh modes, equation (\ref{GeneralUnruh}), we find
\begin{eqnarray}
\hat c_{\omega}^{\prime} &=& \hat c_{\omega}^{s} + \alpha g^*(\omega) L_{\omega}, \nonumber\\
\hat d_{\omega}^{\prime} &=& \hat d_{\omega}^{s} + \alpha^* g(\omega) M_{\omega},
\end{eqnarray}
where $L_{\omega} = \cosh r_{\omega} - z \sinh r_{\omega}$, $M_{\omega} = z \cosh r_{\omega} - \sinh r_{\omega}$, the operators $ \hat c_{\omega}^{s}$ and 
$\hat d_{\omega}^{s}$ are the output Unruh operators in the absence of displacements. For a uniformly accelerated single-mode squeezer, 
\begin{eqnarray}\label{IOUnruh:squeezer}
\hat{c}^{s}_{\omega} &=& \hat{c}_{\omega} + g^*(\omega) \cosh r_{\omega} \big[ \hat{b}^R_g(\cosh r -1)  + \hat{b}^{R\dag}_g \sinh r \big], \nonumber\\
\hat{d}^{s}_{\omega} &=& \hat{d}_{\omega} - g(\omega) \sinh r_{\omega} \big[ \hat{b}^{R \dag}_g(\cosh r -1)  + \hat{b}^{R}_g \sinh r \big].
\end{eqnarray}
The total Minkowski particle number operator 
\begin{eqnarray}\label{PN:twoD}
\hat N = \int \mathrm{d} \omega ~ (\hat c_{\omega}^{\prime\dag} \hat c_{\omega}^{\prime} + \hat d_{\omega}^{\prime\dag} \hat d_{\omega}^{\prime})
= |\alpha|^2 \int \mathrm{d} \omega ~ |g(\omega)|^2 (L_{\omega}^2 + M_{\omega}^2) + (\hat n_c + \hat n_d) + \hat Y_c + \hat Y_d,
\end{eqnarray}
where $\hat n_c$ and $\hat n_d$ are the output Unruh particle numbers in the absence of displacements,
\begin{eqnarray}
\hat n_c = \int \mathrm{d} \omega ~ \hat{c}^{s\dag}_{\omega} \hat{c}^{s}_{\omega}, ~~~~~~ \hat n_d = \int \mathrm{d} \omega ~ \hat{d}^{s\dag}_{\omega} \hat{d}^{s}_{\omega};
\end{eqnarray}
$\hat Y_c$ and $\hat Y_d$ are defined as
\begin{eqnarray}\label{defintion:ycyd}
\hat Y_c &=& \alpha^* \int \mathrm{d} \omega ~ g(\omega) L_{\omega} \hat{c}^{s}_{\omega} + \alpha \int \mathrm{d} \omega ~ g^*(\omega) L_{\omega} \hat{c}^{s\dag}_{\omega}, 
\nonumber\\
\hat Y_d &=& \alpha \int \mathrm{d} \omega ~ g^*(\omega) M_{\omega} \hat{d}^{s}_{\omega} + \alpha^* \int \mathrm{d} \omega ~ g(\omega) M_{\omega} \hat{d}^{s\dag}_{\omega}.
\end{eqnarray}
To second order of $\alpha$, the variance of the total particle number is 
\begin{eqnarray}
\langle 0 | \hat N^2 | 0 \rangle - \langle 0 | \hat N | 0 \rangle^2 = \langle 0 | \hat Y_c^2 | 0 \rangle + \langle 0 | \hat Y_d^2 | 0 \rangle + 2 \langle 0 | \hat Y_c \hat Y_d | 0 \rangle.
\end{eqnarray}
For a uniformly accelerated single-mode squeezer, the expectation values $\langle 0 | \hat Y_c^2 | 0 \rangle$, $\langle 0 | \hat Y_d^2 | 0 \rangle$ and
$\langle 0 | \hat Y_c \hat Y_d | 0 \rangle$ can be calculated straightforwardly from equations (\ref{defintion:ycyd}) and (\ref{IOUnruh:squeezer}). 
\begin{eqnarray}
\langle 0 | \hat Y_c^2 | 0 \rangle &=& |\alpha|^2 (\mathcal{I}_c - 2 z \mathcal{I}_{cs} + z^2 \mathcal{I}_s) 
+ 2  |\alpha|^2\big[E_c + E_{cc} \cos(2\phi)\big] (\mathcal{I}_c - z \mathcal{I}_s)^2, \nonumber\\
\langle 0 | \hat Y_d^2 | 0 \rangle &=& |\alpha|^2 (z^2 \mathcal{I}_c - 2 z \mathcal{I}_{cs} + \mathcal{I}_s) 
+ 2  |\alpha|^2\big[E_d + E_{dd} \cos(2\phi)\big] (z \mathcal{I}_c - \mathcal{I}_s)^2, \nonumber\\
\langle 0 | \hat Y_c \hat Y_d | 0 \rangle &=& 2  |\alpha|^2\big[E_{cd} + \bar E_{cd} \cos(2\phi)\big] (\mathcal{I}_c - z \mathcal{I}_s) (z \mathcal{I}_c - \mathcal{I}_s),
\end{eqnarray}
where the new integral 
$\mathcal{I}_{cs}$ is defined as 
\begin{eqnarray}
\mathcal{I}_{cs} = \int \mathrm{d} \omega ~ |g(\omega)|^2 \cosh r_{\omega} \sinh r_{\omega}.
\end{eqnarray}
Therefore the variance of the quadrature amplitude is 
\begin{eqnarray}\label{Variance:2D}
V(\phi) &=& \frac{\langle 0 | \hat N^2 | 0 \rangle - \langle 0 | \hat N | 0 \rangle^2 }{\langle 0 | \hat N_0 | 0 \rangle} \nonumber\\
&=& 1 + 2 \big\{ \big[E_{cc} (\mathcal{I}_c - z \mathcal{I}_s)^2 + E_{dd} (z \mathcal{I}_c - \mathcal{I}_s)^2 + 2 \bar E_{cd}(\mathcal{I}_c - z \mathcal{I}_s) (z \mathcal{I}_c - \mathcal{I}_s) \big]\cos(2\phi)
\nonumber\\
&& \big[E_c (\mathcal{I}_c - z \mathcal{I}_s)^2 + E_d (z \mathcal{I}_c - \mathcal{I}_s)^2 + 2 E_{cd}(\mathcal{I}_c - z \mathcal{I}_s) (z \mathcal{I}_c - \mathcal{I}_s) \big]
\big\}/{ \big[(1+z^2)(\mathcal{I}_c + \mathcal{I}_s) - 4 z \mathcal{I}_{cs} \big]}, 
\end{eqnarray}
where $\hat N_0$ can be obtained from equation (\ref{PN:twoD}) by omitting $\hat n_c$ and $\hat n_d$. Notice that when $z = 0$ we regain equation (\ref{Variance}) as expected (see also equation (\ref{Nsquared})).

In the narrow bandwidth limit, namely, the central frequency $\omega_0$ of $g(\omega)$ is much greater than the bandwidth $\sigma$, we can approximate the integrals as
\begin{eqnarray}
\mathcal{I}_c \approx \cosh^2 r_{\omega_0}, ~~~~~~
\mathcal{I}_s \approx \sinh^2 r_{\omega_0}, ~~~~~~
\mathcal{I}_{cs} \approx \cosh r_{\omega_0} \sinh r_{\omega_0}.
\end{eqnarray}
By substituting these approximated integrals into equation (\ref{Variance:2D}), it is straightforward to show that when 
\begin{eqnarray}
z = \frac{2\sqrt{\mathcal{I}_c(\mathcal{I}_c - 1)}}{2\mathcal{I}_c-1},
\end{eqnarray}
the variance of the quadrature amplitude is
\begin{eqnarray}
V(\phi) = \cosh(2r) + \sinh(2r) \cos(2\phi).
\end{eqnarray}
This implies $V_{\text{max}} V_{\text{min}} = 1$, indicating that the detected state is pure. 

More generally it can be shown that for an arbitrary wave packet, $g(\omega)$, there exists a non-zero $z$ such that $V_{\text{max}} V_{\text{min}} = 1$. In fact, one can solve the equation
$V_{\text{max}} V_{\text{min}} = 1$ for $z$ based on the variance of the quadrature amplitude equation (\ref{Variance:2D}). Therefore, by appropriately creating a particular additional local
oscillator in the left Rindler wedge, but only by doing so, the inertial detector will see a pure state.

\end{document}